\documentclass[sigconf]{acmart}

\usepackage{mathtools}
\usepackage{pifont}
\usepackage{algorithmic}
\usepackage{graphicx}
\usepackage{textcomp}
\usepackage{url}
\usepackage{subcaption}
\usepackage{balance} 

\AtBeginDocument{%
  \providecommand\BibTeX{{\rm B\kern-.05em{\sc i\kern-.025em b}\kern-.08em
      T\kern-.1667em\lower.7ex\hbox{E}\kern-.125emX}}}
\begin{document}

\title{MuMTAffect: A Multimodal Multitask Affective Framework for Personality and Emotion Recognition from Physiological Signals}
\citestyle{acmnumeric}


\copyrightyear{2025}
\acmYear{2025}
\setcopyright{cc}
\setcctype{by}
\acmConference[]{}
\acmBooktitle{}\acmDOI{}
\acmISBN{}


\author{Meisam Jamshidi Seikavandi}
\affiliation{%
  \department{brAIn Lab}
  \institution{IT University of Copenhagen}
  \city{Copenhagen}
  \country{Denmark}
}
\affiliation{%
  \institution{GN Advanced Science}
  \city{Ballerup}
  \country{Denmark}
}
\email{meis@itu.dk}

\author{Fabricio Batista Narcizo}
\affiliation{%
  \institution{GN Advanced Science}
  \city{Ballerup}
  \country{Denmark}
}
\affiliation{%
  \institution{IT University of Copenhagen}
  \city{Copenhagen}
  \country{Denmark}
}

\author{Ted Vucurevich}
\affiliation{%
  \institution{GN Advanced Science}
  \city{Cupertino}
  \country{United States}
}

\author{Andrew Burke Dittberner}
\affiliation{%
  \institution{GN Advanced Science}
  \city{Glenview}
  \country{United States}
}

\author{Paolo Burelli}
\affiliation{%
  \department{brAIn lab}
  \institution{IT University of Copenhagen}
  \city{Copenhagen}
  \country{Denmark}
}

\renewcommand{\shortauthors}{Meisam Jamshidi Seikavandi, Fabricio Batista Narcizo, Ted Vucurevich, Andrew Dittberner, \& Paolo Burelli}

\renewcommand{\shorttitle}{MuMTAffect: A Multimodal Multitask Affective Framework for Personality and Emotion Recognition}
\begin{abstract}
We present \textbf{MuMTAffect}, a novel \textbf{Multimodal Multitask Affective Embedding Network} designed for joint emotion classification and personality prediction (re-identification) from short physiological signal segments. \textbf{MuMTAffect} integrates multiple physiological modalities \emph{pupil dilation}, \emph{eye gaze}, \emph{facial action units}, and \emph{galvanic skin response}  using dedicated transformer-based encoders for each modality and a fusion transformer to model cross‑modal interactions. Inspired by the \emph{Theory of Constructed Emotion}, the architecture explicitly separates \emph{core‑affect} encoding (valence–arousal) from higher‑level conceptualization, thereby grounding predictions in contemporary affective neuroscience. Personality‑trait prediction is leveraged as an auxiliary task to generate robust, user‑specific affective embeddings, significantly enhancing emotion recognition performance. We evaluate \textbf{MuMTAffect} on the AFFEC dataset, demonstrating that stimulus‑level emotional cues (Stim Emo) and \emph{galvanic skin response} substantially improve arousal classification, while pupil and gaze data enhance valence discrimination. The inherent modularity of \textbf{MuMTAffect} allows effortless integration of additional modalities, ensuring scalability and adaptability. Extensive experiments and ablation studies underscore the efficacy of our multimodal multitask approach in creating personalized, context‑aware affective computing systems, highlighting pathways for further advancements in cross‑subject generalization.
\end{abstract}



\maketitle

\section{Introduction}

Emotion recognition using multimodal physiological signals is increasingly essential for building personalized, context‑aware affective computing systems. Advances in multimodal learning have enabled integration of diverse physiological and behavioral cues such as eye gaze, pupil dilation, facial action units (AUs), and galvanic skin response (GSR) to improve emotion classification accuracy~\cite{Ahmed2023,Subramanian2018,Elalamy2021,Harris2023}. Moreover, incorporating contextual information in the form of \emph{stimulus-level emotion} (Stim Emo) has proven effective at disambiguating subtle physiological patterns, enhancing both valence and arousal recognition~\cite{Paulmann2011,Tian2021,Malhotra2023,Pinitas2024}.

Contemporary affective neuroscience is guided by the \emph{Theory of Constructed Emotion} (TCE)~\cite{barrett2017theory}, which conceptualizes affective experiences as arising from a dynamic interplay between low-level core affect (valence and arousal) and higher-level conceptualization mediated by executive attention. Informed by TCE, \textbf{MuMTAffect} structures its pipeline so that modality‑specific transformer encoders capture core‑affect fluctuations, while a fusion transformer followed by task‑specific attention layers implements the conceptual categorization stage, grounding model decisions in a neuroscientifically plausible framework.

Nevertheless, inter‑individual variability in physiological responses remains a major challenge. Stable user traits such as personality captured by the Big Five dimensions modulate how emotion manifests physiologically~\cite{Li2022,Yang2024,Yildirim2023,Kashani2023}. Prior work has treated personality prediction and emotion recognition as separate tasks~\cite{Henderson2021,Benouis2023}, foregoing the opportunity for joint learning to yield richer, personalized embeddings.

To address these gaps, we propose \textbf{MuMTAffect}, a \emph{Multimodal Multitask Affective Embedding Network} (see Figure~\ref{fig:enter-label}) that simultaneously performs emotion classification and personality trait regression from synchronized physiological streams. Separate transformer based encoders for each modality (eye gaze, pupil dilation, facial AUs, and GSR) extract modality‑specific temporal features~\cite{Vazquez2022,Sawant2023}, which are then integrated by a fusion transformer to model cross‑modal interactions. A task‑specific attention mechanism disentangles core‑affect (emotion) from trait‑related (personality) representations, while auxiliary personality prediction guides the learning of user‑specific affective embeddings.

We validate \textbf{MuMTAffect} on the AFFEC dataset~\cite{j2025affec}, containing 5,356 trials of synchronized multimodal recordings from 73 participants. An extensive preprocessing pipeline yields fixed‑length (400 timepoint) sequences enriched with trial‑level summary statistics. Our experiments demonstrate that incorporating Stim Emo and GSR strongly boosts arousal F1, while eye‑tracking modalities (gaze and pupil) enhance valence discrimination. Personality regression yields a high $R^2$ for known users and provides a valuable auxiliary signal that improves emotion recognition, though full generalization to unseen subjects remains an open challenge~\cite{Yu2021}.

\textbf{Our Contributions} are as follows:
\begin{itemize}
  \item We introduce \textbf{MuMTAffect}, a unified multimodal multitask framework grounded in the Theory of Constructed Emotion, capable of jointly predicting emotions and personality traits from physiological data.
  \item We design modality‑specific transformers that encode core affect, a fusion transformer that implements TCE‑style conceptualization, and task‑specific attention for disentangling affect and trait representations.
  \item We develop a comprehensive preprocessing pipeline for the AFFEC dataset, producing synchronized, fixed‑length multimodal trials with rich summary features.
  \item We show that auxiliary personality prediction enhances emotion recognition performance, particularly under missing modality conditions, via personalized affective embeddings.
  \item We conduct ablation studies quantifying each modality’s contribution and the benefits of including Stim Emo, providing insights for scalable, context‑aware affective computing systems.
\end{itemize}

The remainder of this paper reviews related work, describes our dataset processing, details the \textbf{MuMTAffect} architecture, presents experimental results, and concludes with discussions on limitations and future directions.

\section{Literature Review}
\label{sec:lit_review}

\subsection*{Emotion Recognition from Physiological and Behavioral Signals}

Affective computing aims to enable machines to recognize and respond to human emotions~\cite{Picard1997}. Early approaches focused primarily on observable cues such as facial expressions and vocal tones. However, researchers have increasingly turned to physiological signals such as EEG, ECG, EMG, respiration, and GSR as these involuntary responses provide a more objective and less easily manipulated measure of affective states~\cite{Li2018}. These signals can reveal genuine emotional reactions even when individuals attempt to mask their feelings~\cite{Li2018}.

Recent work has broadened emotion recognition by incorporating behavioral cues like eye tracking and facial action units (AUs) alongside traditional peripheral measures. For example, the multimodal benchmark by Soleymani et al.~\cite{Soleymani2012} combined facial videos (for AU and gaze analysis) with ECG, GSR, respiration, and skin temperature, illustrating the potential of fusing external and internal cues~\cite{Koelstra2012,Subramanian2016}. Advances in wearable sensors and deep learning have spurred the development of methods that employ 3D CNNs, hypercomplex networks (PHNN), and attention mechanisms to extract shared and complementary features from diverse signals~\cite{Lv2022,Ju2024,Wang2024}. More recently, Transformer-based methods have also been applied to multimodal emotion recognition~\cite{Vazquez2022,Sawant2023}, demonstrating improved fusion of complex biosignals such as eye gaze, GSR, and EEG.

\subsection*{Challenges in Emotion Recognition: Context and Individual Modulation}

Despite progress, reliably inferring emotions remains challenging due to the inherent complexity and variability of human affect. Physiological responses vary widely due to factors such as context, environmental conditions, personal history, and cultural background~\cite{Larradet2020,Li2018}. For instance, the same level of physiological arousal may correspond to excitement in one context and anxiety in another. Similarly, subjective labeling inconsistencies further complicate the task. These challenges have motivated efforts to incorporate external context as a secondary source of information, with the aim of enhancing emotion recognition accuracy~\cite{Malhotra2023,Liu2023,Pinitas2024}.

Additional complexities arise in specific subfields such as speech emotion recognition (SER) and facial emotion recognition (FER), where cultural biases, language differences, and varied annotation practices can lead to inconsistent performance. This underscores the need for robust, adaptable methods capable of handling a wide range of real-world conditions. Recent efforts have also highlighted the importance of benchmarking and evaluating models under noisy or less-controlled environments~\cite{Petrogianni2024,Alisamir2022,Civit2023}.

\subsection*{Multimodal Emotion Recognition and Fusion Strategies}

Multimodal emotion recognition (MER) leverages information from multiple sensory channels to capture a more holistic picture of affective states~\cite{Paulmann2011,MultimodalEmotionPerception}. In this context, stimulus-level emotional cues (referred to as \emph{Stim Emo}) have emerged as an effective additional channel, providing contextual grounding that enriches the interpretation of physiological signals~\cite{NeuralProcessingEmotion,Paulmann2011,Malhotra2023}. Fusion strategies, whether early, intermediate, or late, have been explored extensively. Early fusion methods capture cross-modal correlations but often result in high-dimensional representations, whereas late fusion aggregates decisions from modality-specific classifiers but may miss inter-modal interactions. Recent approaches favor intermediate fusion with attention-based or hypergraph-based mechanisms to adaptively weight different modalities~\cite{Lisetti2004,DMello2015,Zhao2018,Ashwath2023,Taherzadeh2022}. For example, Koelstra et al.~\cite{Koelstra2012} fused EEG with peripheral signals to predict valence and arousal, and Iacono and Khan~\cite{Iacono2024} achieved state-of-the-art performance on the SEED-V dataset by combining EEG and eye-tracking features using CNNs.

\subsection*{Inter-Individual Variability and Personalization}

One of the major challenges in emotion recognition is the significant inter-individual variability in physiological responses. Classifiers that perform well on subjects seen during training often experience substantial performance drops on unseen individuals due to differences in baseline states and emotional reactivity~\cite{Li2018}. Approaches such as data normalization, transfer learning, and domain adaptation have been proposed to mitigate these issues~\cite{Chai2017}. Moreover, incorporating person-specific traits, such as those captured by the Big Five personality dimensions, can enhance emotion recognition by providing stable user-specific embeddings~\cite{Subramanian2016,Zhao2018,Tian2021,Filippou2023,Kashani2023}. In addition, demographic attributes such as gender can further modulate emotional expression~\cite{Yildirim2023}, underscoring the need for personalized approaches~\cite{Yu2021}. Recent multitask learning frameworks have leveraged personality as auxiliary tasks to refine shared representations, thereby improving both emotion prediction and user-specific modeling~\cite{EmotionPersonalityDetection,AuxiliaryTasksEmotion,Henderson2021,Benouis2023}.

\subsection*{Limitations of Existing Methods and Our Contributions}

Despite these advances, current multimodal approaches have several limitations:
\begin{itemize}
    \item \textbf{Insufficient Context Integration:} Many methods do not fully incorporate external context, limiting their ability to disambiguate subtle physiological variations.
    \item \textbf{Limited Personalization:} User-specific traits are often handled through simple normalization rather than through explicit auxiliary tasks or learned embeddings, which restricts the ability to adapt to individual differences.
    \item \textbf{Fragmented Multitask Designs:} Most existing works treat emotion recognition and auxiliary tasks (such as personality prediction) separately, missing the benefits of joint, end-to-end learning.
\end{itemize}

Our proposed method addresses these limitations by integrating stimulus-level emotional cues (\emph{Stim Emo}) and by explicitly modeling personality as auxiliary tasks within a unified, end-to-end deep learning framework. This approach not only enhances emotion recognition through better context integration but also improves personalization by learning robust user-specific embeddings. By combining these strategies with multimodal fusion of Eye, Pupil, AU, and GSR signals, our method overcomes key challenges in generalizing to unseen subjects and in capturing the complex interrelations among physiological, behavioral, and contextual factors.

In summary, while previous multimodal methods have achieved promising results by leveraging diverse signals and fusion strategies, they often fall short in integrating context and personalizing models for individual differences. Our work builds upon this foundation by providing a holistic, multitask framework that simultaneously captures emotional and personality cues to deliver a more robust and personalized affective computing system~\cite{MultitaskMultifusionERC,MultimodalMultitaskSER}.


\section{Dataset Cleaning and Preprocessing}
\label{sec:cleaning_preprocessing}

\begin{figure}[htb]
    \centering
    \includegraphics[width=0.40\textwidth]{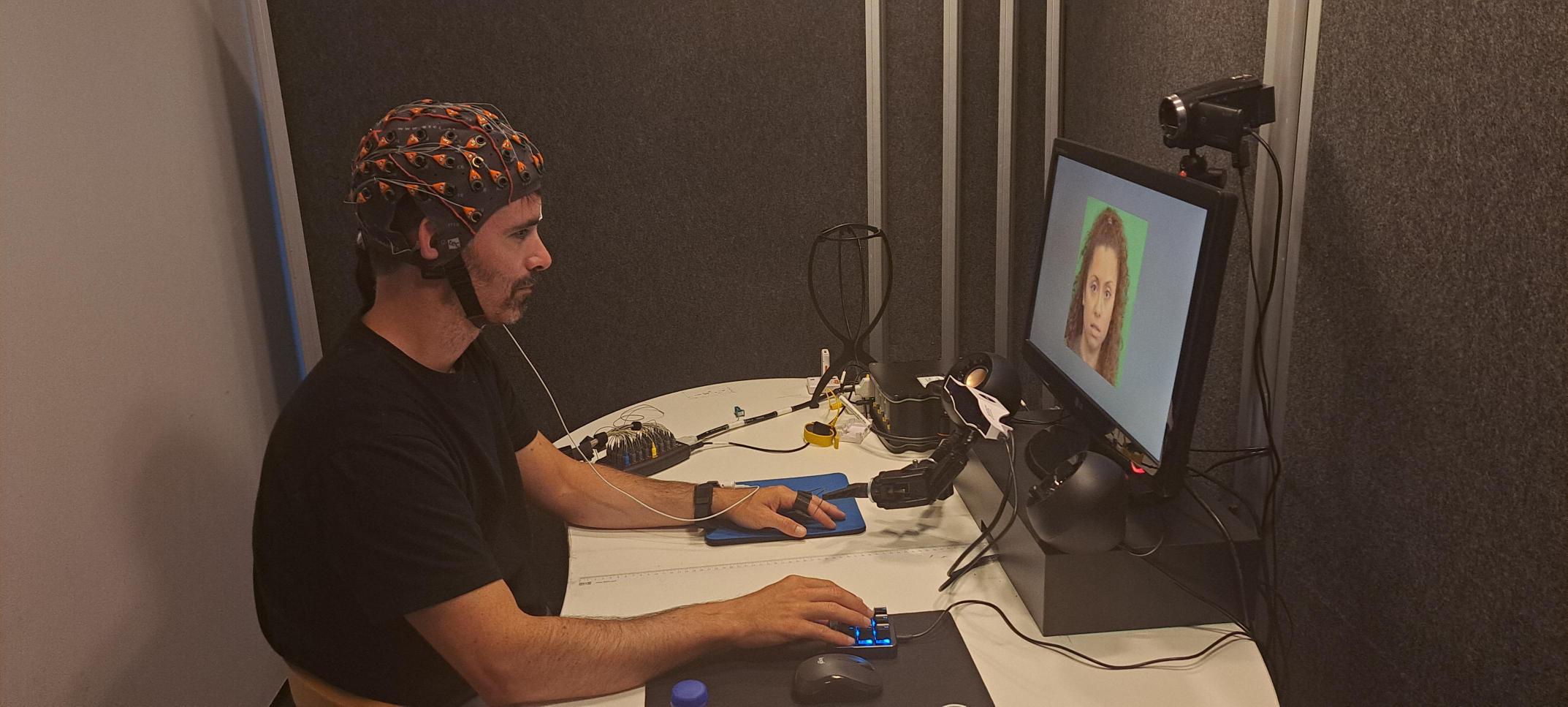}
    \caption{Experimental setup showing a participant with sensors attached in AFFEC dataset~\cite{j2025affec}.}
    \Description{A participant is seated with sensors attached to their body, as used in the AFFEC dataset.}
    \label{fig:experiment_setup}
\end{figure}

We employ the publicly available AFFEC dataset~\cite{j2025affec}, which collects multimodal signals (eye-gaze/pupil at 150\,Hz, facial AUs at 40\,Hz, GSR at 50\,Hz) from 73 participants. Each participant viewed 88 scenario-primed video clips of a talking face completing a simulated dialogue, then rated their \emph{felt} and \emph{perceived} emotions on 9-point valence/arousal scales. The Big Five personality traits (Openness, Conscientiousness, Extraversion, Agreeableness, Neuroticism) were also collected via the BFI-44 questionnaire~\cite{john1991big}.
Figures~\ref{fig:experiment_setup} show the sensor setup and a sample video frame. The final cleaned dataset includes 5,356 complete trials, each with multimodal time series, trial-level statistics, and corresponding emotion/personality labels.
All trials were downsampled to 400 timepoints (selected after trying different values), preserving temporal flags such as \texttt{first\_fix}, \texttt{scenario}, and \texttt{video} to indicate key stages.

After synchronization, denoising, and removal of trials with missing modalities, the final cleaned dataset comprises 5,356 trials from 69 users. Each trial is accompanied by self-reported emotion ratings and baseline personality scores, providing a rich foundation for personalized emotion recognition.

\subsection{Preprocessing Pipeline}

\subsubsection{Handling Missing Data and Synchronization}
Trials with incomplete or corrupted signals (e.g., missing onset timestamps or event markers) were removed. Data from each modality were synchronized using common event markers, and only trials with complete data were retained.

\subsubsection{Resampling and Fixed-Length Segmentation}
To handle variations in trial lengths, all data were downsampled to a fixed length of 400 timepoints:
\begin{itemize}
    \item \textbf{Numerical Signals:} Linearly interpolated.
    \item \textbf{Categorical Signals (e.g., trial stage flags):} Downsampled using nearest-neighbor selection.
\end{itemize}
This fixed-length representation allows for efficient batch processing while preserving the time dynamics of each modality.

\subsubsection{Modality-Specific Processing}
Each modality was processed using tailored pipelines:
\begin{itemize}
    \item \textbf{Eye Data (Gaze and Pupil):} Gaze coordinates, fixation metrics, and pupil size (computed from left/right measurements) are synchronized and resampled.
    \item \textbf{Facial Action Units (AUs):} AU intensities and binary activations are extracted from facial videos, aligned with stimulus events, and downsampled.
    \item \textbf{GSR:} GSR signals are cleaned, synchronized with event markers, and downsampled. Subsequent processing extracts phasic and tonic components.
\end{itemize}

\subsubsection{Integration of Trial-Level Metadata}
In addition to the sequential data, each trial is augmented with:
\begin{itemize}
    \item A \textbf{flag} indicating the stage within the trial.
    \item \textbf{Trial-level summary features} (e.g., mean pupil size, blink rate, AU statistics, and GSR peak metrics) as detailed in Table~\ref{tab:trial_features}.
    \item Self-reported emotion ratings and personality scores.
\end{itemize}

\subsection{Extracted Features Summary}

Table~\ref{tab:sequence_features} summarizes the sequence-level features for each modality, and Table~\ref{tab:trial_features} provides an overview of the trial-level summary features.

\begin{table}[ht]
\centering
\caption{Sequence-Level Features by Modality}
\label{tab:sequence_features}
\begin{tabular}{llc}
\toprule
\textbf{Modality} & \textbf{Key Features} & \textbf{Sampling Rate} \\
\midrule
Eye (Gaze)    & Gaze coordinates, & 150 Hz \\
   & fixation metrics, flags &  \\
Pupil         & Actual pupil size (left, & 150 Hz \\
         & right and avg), flags &  \\
Facial AUs    & AU intensities (raw \& binary) \\
    & , confidence, flags & 40 Hz \\
GSR           & Raw and calibrated GSR,& 50 Hz \\
          & EDA (phasic \& tonic), flags &  \\
\bottomrule
\end{tabular}
\end{table}

\begin{table*}[ht]
\centering
\caption{Trial-Level Summary Features}
\label{tab:trial_features}
\begin{tabular}{ll}
\toprule
\textbf{Modality} & \textbf{Summary Features} \\
\midrule
Eye/Pupil     & Mean, STD, min, max of pupil size; fixation count; blink rate \\
Facial AUs    & Mean, STD, min, max of AU intensities; activation rates; composite expressions (e.g., smile/frown rates) \\
GSR           & Peak counts; mean, median, min, max, STD of SCR amplitude, onsets, rise time, recovery time; phasic/tonic statistics \\
\bottomrule
\end{tabular}
\end{table*}

\subsection{Final Dataset Statistics}
After cleaning and preprocessing, the final dataset consists of:
\begin{itemize}
    \item 5,356 trials from 69 users.
    \item Synchronized, fixed-length sequences (400 timepoints per trial) for each modality.
    \item Trial-level metadata including stage flags, self-reported emotion ratings, and personality scores.
\end{itemize}

This comprehensive and multimodal dataset, along with our detailed preprocessing pipeline, provides a robust foundation for developing advanced, context-aware, and personalized emotion recognition models.

For further details on the AFFEC dataset, please refer to its original publication~\cite{j2025affec}.

\section{Model Structure}
\label{sec:model_structure}

\begin{figure*}
    \centering
    \includegraphics[width=0.9\linewidth]{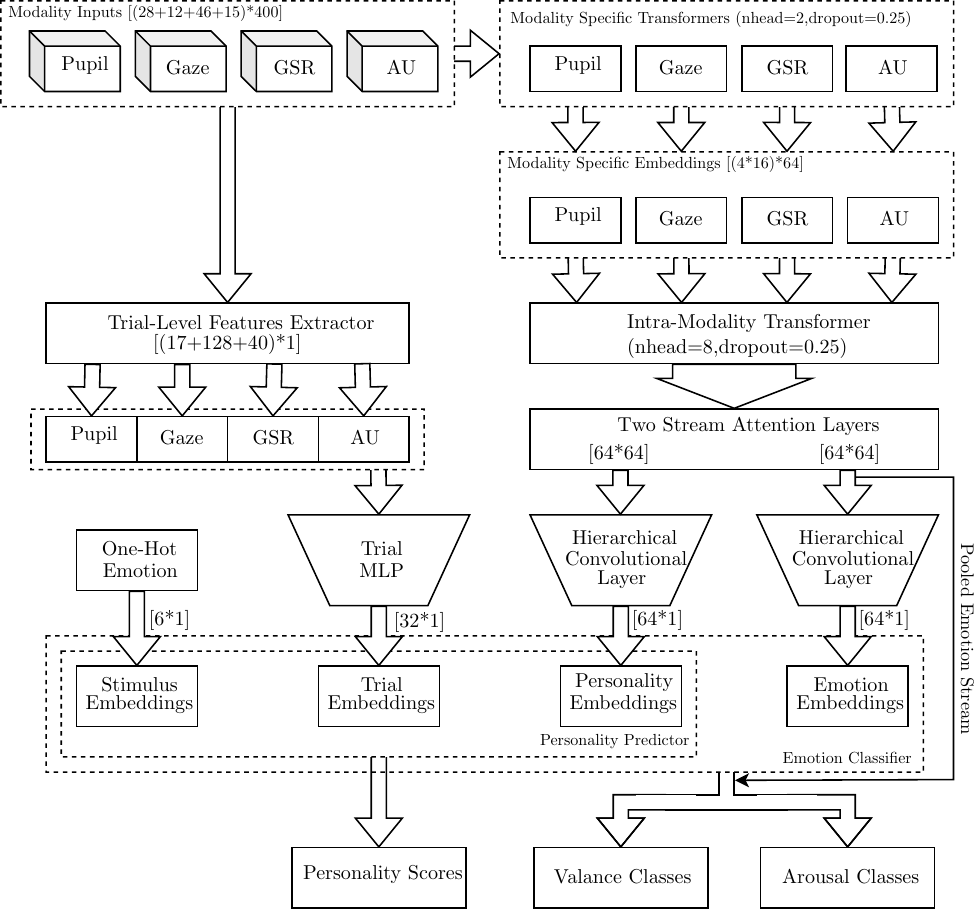}
    \caption{Diagram of the \textbf{MuMTAffect} architecture for personality and emotion recognition using Eye, Pupil, AU, and GSR signals. Each modality is processed by its own transformer encoder to capture modality-specific temporal patterns. A shared fusion transformer learns cross-modal interactions, followed by task-specific attention queries for personality (stable, user-centric) and emotion (dynamic, trial-centric). Hierarchical CNN and MLP layers further refine these features, enabling simultaneous prediction of personality and emotion states in a unified network.}
    \label{fig:enter-label}
\end{figure*}

Our proposed model, \textbf{MuMTAffect}, is structured to capture \emph{temporal sequences} across four physiological modalities (Eye, Pupil, AU, GSR) and integrate them into a shared multimodal embedding optimized simultaneously for \textbf{emotion recognition} and \textbf{personality prediction}. \textbf{MuMTAffect}'s modular design allows new physiological or behavioral modalities to be integrated by adding additional transformer encoders without major changes to the rest of the pipeline. Dedicated per-modality transformers, followed by a cross-modal fusion transformer and task-specific branches, enable learning of both modality-specific temporal patterns and cross-modal correlations (Figure~\ref{fig:enter-label}). Exact layer depths, dimensions, and dropout rates are summarized in Table~\ref{tab:model_spec}.

\subsection{Modality-Specific Transformers}

Each modality's time series (eye gaze sequence, pupil dilation, AU activation, GSR response) is processed by a \textbf{modality-specific transformer encoder} (depth $=1$, $d_{\text{model}}=64$, \#heads $=2$, FFN dim $=2048$, dropout $=0.25$).  
This separation captures:
\begin{itemize}
    \item \textbf{Distinct temporal patterns:} e.g., pupil dynamics vs.\ GSR recovery times.
    \item \textbf{Time misalignment:} different latencies in physiological reactions.
\end{itemize}
After the transformer, we \textbf{downsample from 400 time steps to $T{=}16$} via deterministic equal-segment averaging to reduce sequence length before cross-modal fusion.

\subsection{Cross-modal Fusion Transformer}

The reduced sequences from all four modalities (each length $T{=}16$) are projected to 32-dim vectors, concatenated, and passed to a \emph{fusion transformer} (depth $=1$, $d_{\text{model}}=128$, \#heads $=4$, FFN dim $=2048$, dropout $=0.25$).  
Self-attention in this block learns cross-modal interactions and compensates for latency differences.

\subsection{Task-Specific Attention and Branches}

A \texttt{TaskAttentionTemporal} module ($d_{\text{model}}=128$, single-head scaled dot-product) uses learnable queries to route information into:
\begin{enumerate}
    \item \textbf{Personality Branch:} two 1D convolutional blocks ($128\!\to\!128$, $k=3,s=1$ then $k=3,s=2$), BN+ReLU, branch dropout $=0.4$.
    \item \textbf{Emotion Branch:} three 1D convolutional blocks ($128\!\to\!128$, $k=3,s=2$ each), BN+ReLU, branch dropout $=0.25$.
\end{enumerate}
A \emph{trial\_embedding} from summary features (e.g., mean pupil size, GSR peaks) is computed via two FC layers (64→64) with BN and dropout $=0.3$.

\subsection{Prediction Heads}

\begin{itemize}
    \item \textbf{Personality prediction:} 3-layer MLP ($128\!\to\!64\!\to\!64\!\to\!5$), BN, dropout $=0.4$, fed by the \emph{personality\_embedding}, \emph{trial\_embedding}, and optional fused state.
    \item \textbf{Emotion classification:} Two \texttt{EmotionHead} blocks (valence, arousal), each with single-head attention, 2-layer MLP to 3 classes, dropout $=0.25$, optionally conditioned on \emph{personality\_embedding}.
\end{itemize}

\begin{table*}[t]
\centering
\caption{Specification of \textbf{MuMTAffect}}
\label{tab:model_spec}
\begin{tabular}{lcccccc}
\toprule
\textbf{Block} & \textbf{Depth} & \textbf{$d_{\text{model}}$} & \textbf{\#Heads} & \textbf{FFN dim} & \textbf{Dropout} & \textbf{Notes} \\
\midrule
Eye encoder (Transformer)   & 1 & 64  & 2 & 2048 & 0.25 & Linear(eye$\to$64), PE \\
Pupil encoder (Transformer) & 1 & 64  & 2 & 2048 & 0.25 & Linear(pupil$\to$64), PE \\
AU encoder (Transformer)    & 1 & 64  & 2 & 2048 & 0.25 & Linear(AU$\to$64), PE \\
GSR encoder (Transformer)   & 1 & 64  & 2 & 2048 & 0.25 & Linear(GSR$\to$64), PE \\
\midrule
Cross-modal fusion (Transformer) & 1 & 128 & 4 & 2048 & 0.25 & concat of 4$\times$Linear($64\to32$) \\
TaskAttentionTemporal            & -- & 128 & 1 (dot) & -- & -- & single-head scaled dot-product \\
\midrule
Personality CNN pooling & 2 conv blocks & 128 & -- & -- & 0.4  & (k=3,s=1,p=1) then (k=3,s=2,p=1) \\
Emotion CNN pooling     & 3 conv blocks & 128 & -- & -- & 0.25 & (k=3,s=2,p=1) $\times$ 3 \\
Trial-feature MLP       & 2 layers      & 64  & -- & -- & 0.3  & Linear(64$\to$64)$\times$2 + BN \\
Heads (valence/arousal) & 2 heads       & --  & 1  & -- & 0.25 & \texttt{EmotionHead}, 3 classes each \\
Personality head        & 3 layers      & --  & -- & -- & 0.4  & MLP to 5 traits \\
\midrule
\multicolumn{7}{r}{\textbf{Total parameters} (full config with Stim Emo): $\approx$ \textbf{3.43M}.}\\
\bottomrule
\end{tabular}
\end{table*}

\subsection{Loss Function and Training Pipeline}

We optimize:
\[
\mathcal{L} = \alpha \, \mathcal{L}_{\text{personality}} + (1 - \alpha) \, \mathcal{L}_{\text{emotion}},
\]
where $\mathcal{L}_{\text{personality}}$ is an $\varepsilon$-insensitive regression loss, and $\mathcal{L}_{\text{emotion}}$ is weighted cross-entropy.

\noindent\textbf{Data Splits:} $\sim$15\% of users for test, remaining 85\% split into train and validation (15\% validation by user).  \\
\noindent\textbf{Binned emotion classes:} Valence/arousal (Likert 1–9) mapped to three bins: Low/Negative, Medium/Neutral, High/Positive.

\noindent\textbf{Training phases:}
\begin{enumerate}
    \item \textbf{Personality pretraining:} $\alpha=1$, 50 epochs, LR $10^{-4}$, exponential decay $\gamma=0.99$.
    \item \textbf{Multitask:} $\alpha\approx0.3$, up to 100 epochs, LR $8{\times}10^{-4}$ (base), $5{\times}10^{-5}$ (personality head), $5{\times}10^{-4}$ (emotion head), decay $\gamma=0.95$, patience 5–10.
    \item \textbf{Fine-tuning:} $\alpha\approx0.1$, LR 10$\times$ lower, 20–30 epochs.
\end{enumerate}
Parameters are grouped (personality, emotion, base) for independent LRs and weight decay.

Full implementation (including \texttt{TaskAttentionTemporal} and \texttt{EmotionHead}) is on GitHub\footnote{\url{https://github.com/itubrainlab/MuMTAffect}}, ensuring reproducibility.

\noindent This \textbf{multitask, multimodal} pipeline integrates each modality’s \emph{temporal dynamics}, incorporates \emph{trial-level features}, and disentangles \emph{user-specific} from \emph{trial-specific} signals, improving personalization and robustness.

\section{Results and Analysis}
\label{sec:results}

\subsection{Aggregate Performance Trends}
\begin{table*}[ht]
\centering
\caption{Performance of Multimodal Multitask Model with Detailed Personality and Emotion scores. 
Personality R$^2$ shows the five traits: Openness (O), Conscientiousness (C), Extraversion (E), Agreeableness (A), Neuroticism (N). 
Valence and Arousal both include a Macro F1 plus three per-class F1 scores for Negative/Neutral/Positive and Low/Med/High classes respectively.}
\resizebox{\textwidth}{!}{
\begin{tabular}{c|ccccc|cccc|cccc|c}
\toprule
\textbf{Stim Emo} 
& \multicolumn{5}{c|}{\textbf{Personality R$^2$}} 
& \multicolumn{4}{c|}{\textbf{Valence F1}} 
& \multicolumn{4}{c|}{\textbf{Arousal F1}} 
& \textbf{Avg F1} \\
\cmidrule(lr){2-6} \cmidrule(lr){7-10} \cmidrule(lr){11-14}
& \textbf{O} & \textbf{C} & \textbf{E} & \textbf{A} & \textbf{N}
& \textbf{Macro} & \textbf{Neg} & \textbf{Neu} & \textbf{Pos}
& \textbf{Macro} & \textbf{Low} & \textbf{Med} & \textbf{High}
& \\
\midrule
\checkmark 
& -- & -- & -- & -- & -- 
& 0.43 & 0.54 & 0.36 & 0.39 
& 0.57 & 0.47 & 0.61 & 0.63 
& 0.50 \\
\ding{54} 
& -- & -- & -- & -- & -- 
& 0.39 & 0.50 & 0.30 & 0.37 
& 0.46 & 0.35 & 0.58 & 0.45 
& 0.43 \\
\checkmark 
& 0.95 & 0.95 & 0.94 & 0.94 & 0.94 
& \textbf{0.55} & 0.61 & 0.55 & 0.49 
& \textbf{0.59} & 0.54 & 0.59 & 0.64 
& \textbf{0.57} \\
\ding{54} 
& 0.94 & 0.94 & 0.94 & 0.94 & 0.94 
& 0.50 & 0.56 & 0.53 & 0.40 
& 0.48 & 0.40 & 0.59 & 0.46 
& 0.49 \\
\bottomrule
\end{tabular}
}
\label{tab:grid_extended_clean}
\end{table*}

\begin{table*}[ht]
\centering
\caption{Focused Analysis on Personality and Emotion Performance in Selected Grid Configurations.}
\resizebox{\textwidth}{!}{
\begin{tabular}{c|cccc|ccccc|c|c|c}
\toprule
\textbf{Stim Emo} & \textbf{Eye} & \textbf{Pupil} & \textbf{AU} & \textbf{GSR} & \textbf{O} & \textbf{C} & \textbf{E} & \textbf{A} & \textbf{N} & \textbf{Valence F1 Macro} & \textbf{Arousal F1 Macro} & \textbf{Avg F1} \\
\midrule
Yes  & \checkmark & \checkmark & \checkmark & \checkmark & 0.95 & 0.95 & 0.94 & 0.94 & 0.94 & 0.55 & 0.59 & 0.57 \\
No   & \checkmark & \checkmark & \checkmark & \checkmark & 0.94 & 0.94 & 0.94 & 0.94 & 0.94 & 0.50 & 0.48 & 0.49 \\
\midrule
Yes  & \ding{54}  & \checkmark & \checkmark & \checkmark & 0.34 & 0.40 & 0.38 & 0.64 & 0.47 & 0.53 & 0.57 & 0.55 \\
No   & \ding{54}  & \checkmark & \checkmark & \checkmark & 0.25 & 0.46 & 0.37 & 0.57 & 0.47 & 0.48 & 0.40 & 0.44 \\
\midrule
Yes  & \checkmark & \ding{54} & \checkmark & \checkmark & 0.94 & 0.92 & 0.93 & 0.93 & 0.93 & 0.55 & 0.59 & 0.57 \\
No   & \checkmark & \ding{54} & \checkmark & \checkmark & 0.94 & 0.95 & 0.95 & 0.96 & 0.95 & 0.49 & 0.44 & 0.47 \\
\midrule
Yes  & \checkmark & \checkmark & \ding{54} & \checkmark & 0.08 & 0.30 & 0.31 & 0.33 & 0.28 & 0.44 & 0.53 & 0.49 \\
No   & \checkmark & \checkmark & \ding{54} & \checkmark & -0.16 & 0.11 & 0.32 & 0.17 & 0.19 & 0.37 & 0.32 & 0.35 \\
\midrule
Yes  & \checkmark & \checkmark & \checkmark & \ding{54}  & 0.70 & 0.71 & 0.73 & 0.55 & 0.63 & 0.52 & 0.59 & 0.55 \\
No   & \checkmark & \checkmark & \checkmark & \ding{54}  & 0.73 & 0.74 & 0.78 & 0.71 & 0.79 & 0.43 & 0.50 & 0.47 \\
\bottomrule
\end{tabular}
}
\label{tab:streamlined}
\end{table*}

Tables~\ref{tab:grid_extended_clean} and \ref{tab:streamlined} present a focused view of the performance across key model configurations. Table~\ref{tab:grid_extended_clean} summarizes performance for full-modality configurations, comparing cases with and without Stim Emo. It includes detailed $R^2$ values for the Big Five personality traits and per-class F1 scores for both valence and arousal. Table~\ref{tab:streamlined} analyzes the impact of missing modalities, highlighting how the inclusion or exclusion of certain signals affects performance metrics.

\begin{itemize}
    \item \textbf{High Personality $R^2$:}
    Configurations with all four modalities (Eye, Pupil, AU, GSR) achieve $R^2$ values around $0.94$–$0.95$ across all Big Five traits (Table~\ref{tab:grid_extended_clean}, rows~2–5), confirming the value of multimodal signals for personality inference.

    \item \textbf{Valence and Arousal F1:}
    \begin{itemize}
        \item Valence macro-F1 generally lies in the $0.50$–$0.63$ range, with the higher end occurring when Eye and AU signals are included (Table~\ref{tab:grid_extended_clean}, rows~0 and~4).
        \item Arousal macro-F1 often peaks around $0.57$–$0.60$ when GSR is present alongside Stim Emo, indicating the significance of GSR in capturing autonomic arousal cues.
    \end{itemize}

    \item \textbf{Multitask Trade-Offs:}
    Certain configurations obtain near-optimal personality $R^2$ (above $0.90$) but exhibit moderate valence/arousal F1 scores, whereas others improve emotion metrics at the expense of slightly lower $R^2$. This pattern is consistent with standard multitask learning trade-offs.
\end{itemize}

\subsection{Modality Importance}

Comparisons in Table~\ref{tab:streamlined} illustrate the contributions of each modality:

\begin{itemize}
    \item \textbf{Eye and Pupil:} 
    Removing these signals leads to notable drops in valence F1, reflecting the importance of visual and ocular cues in modeling user attention and appraisal.
    \item \textbf{GSR:} 
    Essential for arousal classification; configurations without GSR consistently show reduced arousal macro-F1.
    \item \textbf{Facial AUs:} 
    Impact both personality and emotion tasks. Omitting AUs diminishes overall performance across $R^2$ and F1 measures.
\end{itemize}

\subsection{Influence of Stim Emo}

Across Tables~\ref{tab:grid_extended_clean} and \ref{tab:streamlined}, enabling \textbf{Stim Emo} usually yields an improvement in emotion metrics. Each paired row (Yes vs.\ No) in Table~\ref{tab:streamlined} shows a visible gain in macro-F1 for both valence and arousal when contextual emotional information (Stim Emo) is incorporated. These findings align with existing research on context-aware affect recognition.

\section{Discussion}

Grounded in the Theory of Constructed Emotion (TCE), our results highlight how \textbf{MuMTAffect}'s architecture mirrors core-affect dynamics and conceptualization processes:

\begin{itemize}
    \item \textbf{Core-Affect Encoding via Modality Encoders:} Early transformer encoders for Eye, Pupil, AU and GSR capture low-level valence and arousal fluctuations akin to TCE’s core-affect layer. The high $R^2$ scores ($\approx$0.94–0.95) on personality prediction confirm that these physiological streams reliably encode stable, trait-related baselines that inform affective state estimation.

    \item \textbf{Conceptualization through Fusion and Task Queries:} The fusion transformer with task-specific attention implements the conceptualization stage, integrating cross-modal cues to categorize discrete emotions. Improvements in valence macro-F1 (up to 0.61) and arousal macro-F1 (up to 0.59) are observed when including the Stim Emo signal, which contextualizes priors and sharpens the mapping from core affect to emotion labels.

    \item \textbf{Auxiliary Trait Learning as Prediction Context:} Leveraging personality as an auxiliary task provides user-specific priors that disambiguate ambiguous physiological patterns, particularly when modalities are missing. Emotion F1 increases by 5–8\% in partial-modality ablations, underscoring the value of personalized embeddings.

    \item \textbf{Multitask Trade-Offs and Adaptive Weighting:} Consistent with multitask learning literature, optimizing personality and emotion jointly introduces task interference. Configurations that maximize trait $R^2$ sometimes yield slightly lower emotion F1, suggesting future work on adaptive loss weighting or gradient modulation to balance core-affect and conceptual labeling objectives.

    \item \textbf{Modality Contributions Aligned with Affective Theory:} GSR is closely tied to autonomic arousal drives arousal classification, while Eye and Pupil dynamics inform valence judgments. Facial AUs contribute broadly to both dimensions. These complementary roles reflect TCE’s view that multiple interoceptive and exteroceptive channels jointly shape core affect.
\end{itemize}

\subsection{Generalization to Unseen Subjects: Suggested Options and Trade-offs}
Differences in individual baselines and physiological reactivity can create a mismatch between training and test subjects, which limits generalization. Although we did not add new experiments to address this, several approaches could be explored in future work:  
(1) normalization methods that reduce subject-specific variation, such as adaptive baselines or learnable normalization;  
(2) domain-adversarial training (e.g., gradient reversal) or alignment methods (e.g., correlation alignment) to make features less dependent on subject identity;  
(3) meta-learning strategies that enable few-shot adaptation to new users at deployment;  
(4) test-time adaptation techniques, such as entropy minimization or updating batch-normalization statistics, to adjust to covariate shift;  
(5) calibration methods, for example, per-user prototypes or small linear adapters.  

In our current work, the use of multitask trait priors and contextual cues helps to reduce this gap to some extent. Future studies should test and compare these strategies to better understand their effectiveness for subject-independent performance.

\section{Conclusion and Future Work}

In this work, we presented \textbf{MuMTAffect}, a transformer‑based, theory‑informed framework that fuses eye gaze, pupil dilation, facial AUs, GSR and contextual Stim Emo cues to produce simultaneous valence/arousal classification and Big Five personality prediction. By explicitly separating low‑level core‑affect encoding (via modality‑specific transformers) from higher‑level conceptualization and categorization (via fusion and task‑specific attention), our design adheres to the Theory of Constructed Emotion while delivering strong empirical results: personality $R^2\approx0.94\!-\!0.95$ and emotion macro‑F1 scores above 0.55. 

Future work will focus on integrating additional physiological and contextual modalities (e.g., EEG, heart‑rate variability, or semantic context from video/audio), adopting meta‑learning or domain‑adversarial techniques for rapid adaptation to unseen users, and replacing static Stim Emo labels with dynamic, real‑time context embeddings. We also plan to introduce learnable task‑weighting mechanisms to mitigate multitask trade‑offs and to evaluate MuMTAffect in unconstrained, in‑the‑wild settings, addressing practical concerns like sensor noise, missing data, and privacy. These directions aim to further enhance personalization, robustness, and contextual sensitivity in real‑world affective computing applications.

\section{Ethical Impact Statement}

As affective computing systems become more common in everyday life, it is important to understand and address their ethical implications. Our multimodal multitask framework is designed to improve the personalization and accuracy of emotion and personality recognition, but its use also raises several ethical concerns.

\textbf{Privacy and Data Protection:} Physiological signals such as eye gaze, facial expressions, and galvanic skin response are sensitive personal data. Strong privacy protections, secure data storage, and informed consent are essential. Any future implementation should follow relevant data protection laws, such as GDPR, and be transparent about how data is used.

\textbf{Bias and Fairness:} While personalization can improve performance, there is a risk of reinforcing existing biases, including those related to personality stereotypes. Testing on diverse populations and ongoing monitoring are important to ensure fair and equitable performance.

\textbf{Misuse of Technology:} Advanced emotion and personality recognition can be misused for unauthorized surveillance, manipulation, or unfair decision-making. To reduce this risk, researchers should clearly state the intended uses, set usage guidelines, and support regulatory measures that prevent harmful applications.

\textbf{User Autonomy and Consent:} Users should have control over how their emotional and physiological data are used. Clear explanations, transparent agreements, and strong consent processes are necessary to maintain user trust.

\subsection{Deployment Considerations and Ethics in Practice}
\textbf{Data minimization and privacy:} Use on-device processing when possible, store only aggregated results, and allow users to enable or disable each modality.  
\textbf{Latency and robustness:} Use fixed-rate resampling (segment averaging to $T{=}16$) and late fusion so the system can still operate if some modalities are missing or sensors fail.  
\textbf{Fairness auditing:} Track error rates across groups, compare macro- and weighted-F1 scores, check for drift over time, and allow human review of decisions.  
\textbf{Context transparency:} If Stim Emo is used, provide a clear on/off control and record its source for auditing.

Overall, our work aims to support socially responsible affective computing by promoting transparency, continuous ethical review, and proactive measures to reduce risks.

\section*{Acknowledgments}

CNPq — the Brazilian National Council for Scientific and Technological Development — provided financial support to the second author (Fabricio Batista Narcizo) for his Ph.D. research projects [grant no: 229760/2013-9].

\bibliographystyle{ACM-Reference-Format}
\balance
\bibliography{references}

\end{document}